\documentclass[prl,twocolumn,amsmath,amssymb]{revtex4}
\usepackage[T2A]{fontenc}
\usepackage[koi8-r]{inputenc}

\usepackage{epsf,graphicx}
\begin{document}
\title{Study of size effects in the structural transformations in $\textit{bcc}$ Zr films}
\author{E.B. Dolgusheva and V.Yu. Trubitsin}
\date{}
\email{tvynew@otf.pti.udm.ru}
\affiliation{Physical-Technical Institute,
Ural Branch of Russian Academy of Sciences,
132 Kirov Str.,426001 Izhevsk, Russia}
\date{\today}

\begin{abstract}

The effect of the film thickness and temperature on structural phase transformations  in infinite films of bcc zirconium with (001) crystallographic orientation is studied by the molecular dynamics method with a many-body potential of interatomic interaction obtained within the embedded atom model.
It is shown that the mechanism, sequence and final structures of phase transitions in bcc Zr films essentially depend on the film thickness. The films with (001) surface up to 6.1 nm thick experience an orientational transition into the bcc (110) phase through an intermediate metastable fcc phase, and then, on cooling, a diffuse
bcc (110) $\rightarrow$ hcp transition is observed. In films 6.1 to 8.2 nm thick there forms, on cooling, a twin fcc phase as a result of shear deformation, so that the film surface acquires stepped relief. With further increase of the bcc (001) film thickness there occurs martensitic transformation with the formation of a twin hcp structure, and the film has a wavy surface.  
The bulk and elastic moduli are calculated for the bcc and fcc lattices. It is shown that for the fcc phase the lattice stability conditions are satisfied in both bulk and film systems.    
\end{abstract}
\maketitle

The possibility of the formation of anomalous phases in thin films of transition d-metals and the conditions for their stabilization have long been discussed in the literature \cite{Bublik, Komnik,Gladkix}. Such anomalous phases include polymorphous structures not observed in bulk samples, but stabilized in thin films. It is well known, for example, that on the equilibrium P-T phase diagram for the transition metals Ti, Zr, Hf in macroscopic state the fcc phase is lacking. At low temperature and pressure all of these three metals have a hcp structure  ($\alpha$  phase) which with increasing temperature turns into a $\beta$ phase with bcc lattice. 
The high-pressure bcc phase is also realized at room temperature. At low temperatures an increase in pressure leads to a sequence of structural transformations:
hcp $\rightarrow \omega \rightarrow$ bcc. In zirconium, for instance, the  $\omega$ phase exists in a pressure range from 2.2 to 30-35 GPa.
At the same time, it has been experimentally found that in Ti thin films grown on   NaCl, Al, Ni, SiC  substrates a fcc structure is realized \cite{Wawner, Saleh1, Jankowski, Saleh2}. Fcc hafnium was detected in Hf/Fe multilayers \cite{Zhang}. In Ref.\cite{Chopra} it is stated that fcc Zr films thick ~ 500\AA{} were obtained by deposition on glass substrates at $250 - 400^{\circ}C$, with increasing temperature above $ 450^{\circ}C$ a mixture of fcc and hcp structures was observed. In another paper \cite{Hill} fcc zirconium films of 50 monolayers were reported to be obtained in the temperature range 1300K - 1800K.
In Ref.\cite{Manna} were obtained nanoparticles of polycrystalline fcc zirconium with fcc grains of 5-10 nm in diameter and a lattice parameter ranging from 0.4686 nm to 0.4714 nm which was shown to increase with decreasing grain size.  
However, in another experimental work \cite{Prozenko} the anomalous fcc phases of Ti, Zr, and Hf are considered as interstitial impurity phases of residual oxygen atoms. The authors cast some doubt on the possibility of the polymorphous transition realization through the mechanism suggested in Ref.\cite{Bublik}, adducing thermodynamic estimates of the critical film thickness for a phase transition to occur in a free polycrystalline film. They believe the critical thickness of an hcp - fcc transition to be 6.1 nm. Based on an analysis of their own experimental data and those available in the literature, the authors of Ref.\cite{Prozenko} conclude  that "by now (2009) there has been no one \textit{direct} experiment which confirms  the formation of polymorphous modifications not known in bulk samples", and "in  ultrahigh vacuum the formation of only those modifications occurs which are proper to substance in a bulk state".  Thus, various groups of experimenters disagree on the nature of anomalous phases arising in thin films. Some state that the anomalous size-induced phases observed are of polymorphous nature, others consider them as being of impurity character.

Theoretical calculations of the total energy of the Zr ground state within the density functional theory show the fcc structure to be energetically preferable to the bcc one. In Ref. \cite{Xiong} on the basis of first-principles calculations of the Gibbs energy for free nanoparticles of titanium and zirconium, the critical particle size was determined as a function of the temperature at which the fcc lattice becomes stable. At $T=300 K$ this size for Ti and Zr is close to 6 nm, which agrees well with the experimental data \cite{Manna}.  

A possibility for the existence of a metastable fcc phase in zirconium has also been discussed in some works based on the use of the molecular dynamics method. In Refs. \cite{Pinsook, Morris}, for example, the fcc structure was found to form at twin boundaries in the hcp phase of zirconium, and in Refs.
\cite{Ruda,Kucherov} the formation of a fcc ZR phase occurred near the crack tip on the boundaries of twins with hcp structure. A molecular-dynamics study of martensite structures in Zr nanotubes \cite{Thompson} also revealed regions of fcc phase connecting the nearest hcp domains. 
Up to now, there has been no systematic study of the phase size-effect, the conditions of structural stabilization, new phase formation and morphology, mechanisms of phase transformation in nanomaterials of the Zr group.
The aim of this work is to investigate the temperature stabilization region of different crystalline structures in zirconium nanofilms depending on their thickness at a constant pressure (P=0).

\section { Calculation method}

\begin{figure}[tbh]
\begin{center}
\resizebox{0.99\columnwidth}{!}{\includegraphics*[angle=-90]{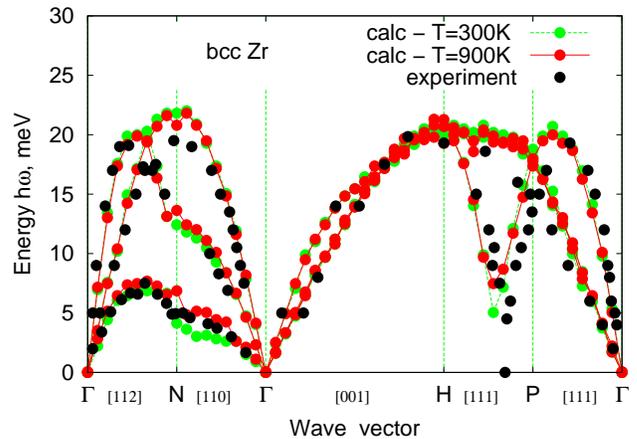}}
\caption{Phonon dispersion curves along the main directions of the Brillouin zone of bcc Zr obtained at $Т=300~K$ and $T=900~K$. Black circles indicate the experimental frequencies\cite{Phonon-bcc}.
 }
\vskip 0.7cm
\label{Fig1}
\end{center}
\end{figure}

The structure stability and physical characteristics of zirconium films were studied by the molecular dynamics (MD) method using the standard XMD package \cite{J-Rifkin}. The many-body potential \cite{Mendelev-Ack} constructed within the "embedded atom" model (EAM) \cite{Daw-Baskes} was chosen to describe interatomic interaction in zirconium. In Ref.\cite{Mendelev-Ack} it is shown that this potential allows one to obtain, to a high degree of accuracy, the bcc and hcp lattice parameters of zirconium, cohesive energy, elastic constants, melting temperature, and other physical characteristics.  The bulk phonon dispersion curves calculated with this potential along the symmetrical directions of the Brillouin zone of bcc zirconium are shown in Fig.\ref {Fig1}. The spectrum was calculated with a Fourier transform of the function of atomic displacement time evolution averaged over an interval of 40 ps. The polarization vectors $\bf {e_{k}}$ with wave vector $\bf {k}$ were calculated in the harmonic approximation as the eigenvalues of dynamical matrix for a bcc lattice with previously determined parameters  corresponding to specified temperatures\cite{our-2009}. A comparison with the experimental data \cite{Phonon-bcc} shows that the potential chosen allows one to reproduce the experimentally observed features of the Zr phonon spectrum and their variation with temperature, including the softening of the transverse N-phonon with decreasing temperature, and so it may be used in simulating the thermal stability of bcc zirconium films.
In this work, the molecular dynamics calculations were performed in two ways: (1) the formation of a crystallite which then is hold, in the state of free evolution, at a specified constant temperature and constant pressure (P=0) for a time no shorter 
 
than $t=100~ps$, and (2) the formation of a crystallite at a high temperature, holding for a time $t=100~ps$, and then a slow cooling at a constant zero pressure with a temperature step $\varDelta T$ and a holding time no lesser than $t = 50~ps$ at each new temperature.  The time step was constant and equal to $\varDelta t = 1 fs$. In what follows the first way will be referred to as "isothermal holding", and the second one as "slow cooling".
Free boundary conditions were specified along the $\textit{z}$ axis. The film thickness on the \textit{z} axis varied from $N_{z}=5$ to $N_{z}=30$ bcc unit cells. In the $\textit{x}$ and $\textit{y }$ directions cyclic conditions were used with  $N_{x,y}$ ($N_x=N_y$) ranging from 7 to 36 u.c. Thus we simulated thin films infinite in $\textit{x}$, $\textit{y}$, with free surfaces on the \textit{z} axis.
 
In all variants, the calculation started with the formation of a crystallite with perfect bcc structure and (001) crystallographic orientation. The lattice parameter was chosen equal to its value in bulk bcc zirconium at the corresponding temperature    \cite{our-2009}. The state of the system was followed up using the atomic radial distribution function (ARDF), the variation of the total energy, and visualization of both the crystallite as a whole and any separate crystallite plane. The belonging of each atom in the system to one of the following structures: bcc, hcp, fcc, and $\omega$ was deduced from its nearest environment.

\section{Size dependence of structural transformations in BCC (001) zirconium films}  
\subsection{Structural transformations under isothermal holding}

The calculations by the isothermal holding technique have shown that for the films with (001) surface there is a critical thickness $N_{z}=17~u.c.$ (about 6.1nm) at which the sequence of structural transformations and their mechanism substantially change. In what follows the films with thicknesses $N_{z}<17~u.c.$  and $N_{z}\geqslant17~u.c.$ will be referred to as "thin" and "thick", respectively. 
\begin{figure}[tbh]
\begin{center}
\resizebox{0.99\columnwidth}{!}{\includegraphics*[angle=-90]{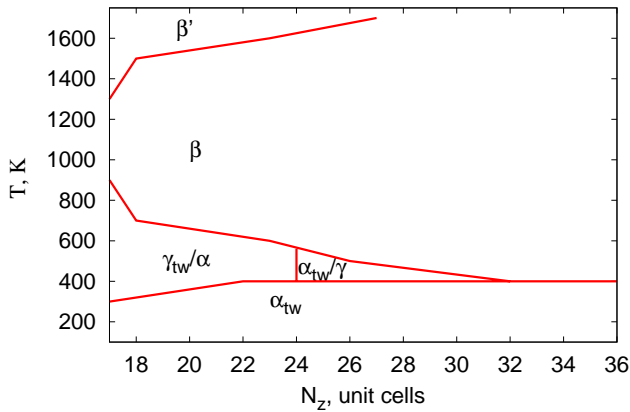}}
\caption{ The $T - N_{z}$ phase diagram obtained by "isothermal holding" for thick bcc (110) Zr films.  $\beta$' is the bcc(110), and $\beta$ is the original bcc(001) phase, $\gamma_{tw}$/$\alpha$ is the twin fcc phase with thin hcp interlayers, $\alpha_{tw}$/$\gamma$ is the twin hcp phase with small fcc inclusions.
 }
\label{Fig2}
\end{center}
\end{figure}
The MD simulation has revealed that bcc (001) thin films are unstable over the whole temperature range considered; they undergo structural transformations the sequence of which is temperature dependent. Under isothermal holding in the range $500K< T\leqslant 1300K$ the sequence of structural transitions is the following: $\beta\rightarrow \gamma \rightarrow \beta '\rightarrow \alpha$. Here $\beta$ is the original film with bcc structure and (001) free surface; $\gamma$ is an intermediate metastable fcc phase; $\beta'$ is a twin bcc structure with (110) free surface, and $\alpha$ is a hexagonal phase.
Besides, the modification of the final phase depends on whether $N_{x,y}$ (the number of cells in the basic crystallite along the x, y axes) is even or odd: at odd values of $N_{x,y}<20$  there forms an orthorhombic $\alpha''$ phase, while at any even $N_{x,y}$  an $\alpha$ phase with ideal HCP lattice arises. In our opinion the $\alpha''$ phase formation is due to the lack of vibrations with wave vectors located  near the point N (${\bf k}=1/2[110]$) of the Brillouin zone of BCC lattice, which are necessary for the $\beta'\to \alpha$ transition to be completed.  This point is discussed in more detail in our previous paper \cite{our-2012}.
At temperatures of thin film formation $T\leqslant\ 500~K$,  the system energy is not sufficient for the $\gamma \to \beta'$  transition which proceeds with an increase in energy, therefore the $\gamma$ phase remains stable. At temperatures above 1300 K the twin bcc structure with (110) surface is stabilized.

In thick films the dependence of structural transformations on the crystallite size along the \textit{x,y} axes is not so significant, and the calculated results may be presented on the Temperature - Film thickness ($T - N_{z}$) phase diagram shown in Fig.\ref{Fig2}. The diagram displays the areas of final stable phases observed in thick films. Note that the orthorhombic phase did not arise in thick films because of the hcp phase being formed here not through the phonon mechanism (cooperative atomic displacement), but as a result of shear deformation.   In addition, in thick films of zirconium the sequence of phase transformations, as well as the temperature ranges of the final phase stability essentially differ from the results obtained for thin films. 
\begin{figure}[tbh]
\begin{center}
\resizebox{0.99\columnwidth}{!}{\includegraphics*[angle=0]{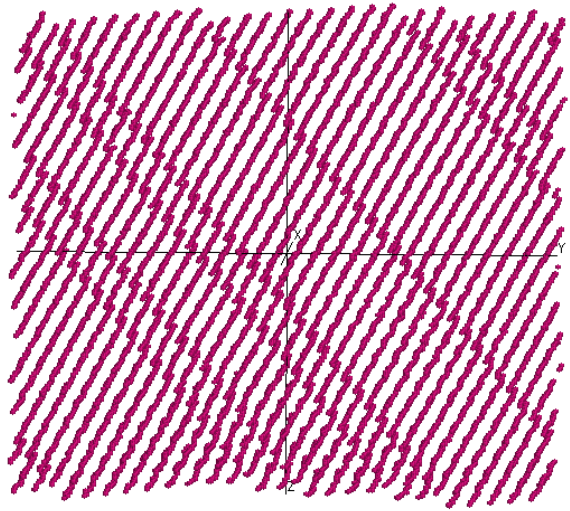}}
\caption{View of the basic crystallite with $N_{z}=17 u.c., N_{x,y}=24 u.c.$ from the [100] direction after isothermal holding at T=400 K.
}
\label{Fig3}
\end{center}
\end{figure}
In thick films reorientation of the original bcc (001) lattice into the bcc (110) one takes place only at temperatures $T\geqslant1400K$. However there is a sufficiently wide temperature range, increasing with the film thickness from [900 K - 1300 K] to [400 K - 1800 K], in which the initial bcc (001) structure remains stable. Simulating bcc (001) thick films by the isothermal holding technique shows that a structural transition from the bcc to a twin fcc phase with a change in the film surface relief occurs in the range of 400 to 700 K. The bcc structure transforms into a twin plate structure where the fcc plates are separated by thin hcp interlayers. The emergence of twins on the free surface of the film results in the formation of a stepped relief, as shown in Fig.\ref{Fig3}. The figure demonstrates a side view of the (001) surface of a film with the basic crystallite dimensions $N_{z}=17~u.c., N_{x,y}=24~u.c.$, at a temperature $T = 400 K$. One can clearly see steps with terraces comprising a few atoms which arise on the surface as a result of shear deformation of fcc layers. In thick films the number of such twin plates increases with $N_{z}$, and consequently increases the portion of the hcp phase.  

An increase in the density of fcc twin plates is also observed with decreasing temperature, i.e. the lower is the temperature of isothermal holding with constant basic crystallite dimensions $N_{z}$ and $ N_{x,y}$, the larger is the portion of hcp phase observed in the final twin structure of a thick film. Moreover, at $T = 200 K$ this kind of twinning may result in the formation of a homogeneous hcp phase which can be considered as a limiting case where each fcc plate involves only two atomic layers, all plates being shifted relative to each other in the [112] direction.   Conversely, the higher is the temperature, the smaller is the number of fcc shearing twins in the crystallite, and the smaller is the hcp phase portion. At certain temperatures, ranging from $500 K$ to $600 K$, in thick films with a thickness   $17~u.c.\leqslant N_{z}\leqslant 19~u.c.$ we observed the formation of a homogeneous fcc phase which remained stable throughout the whole time of simulation. With increasing film thickness the temperature range of the transition to the fcc phase decreases, and also the portion of fcc phase in the final structure gradually decreases. In films with a thickness  $17~u.c.\leqslant N_{z}\leqslant 23~u.c.$ the portion of the fcc phase in the final structure is larger than that of the hcp phase, while in the thickness range  $24~u.c.\leqslant N_{z}\leqslant 31~u.c.$ the portion of the fcc phase is significantly less than the hcp fraction.  From $N_{z} = 32~u.c.$ and on, the fcc phase is not observed at any temperatures,  and only an $\alpha$ phase forms as the result of transformation. Thus, the isothermal holding simulation shows that in thick films of bcc zirconium the phase with fcc lattice arises only in a narrow range of thicknesses and temperatures. Recall that in thin films the fcc phase is stabilized at temperatures below $500~K$ for any value of $N_{x,y}$.

So we came to the conclusion that using the isothermal holding technique which simulates "rapid quenching" of the material, one can obtain a homogeneous fcc film no thicker than a critical value of $17~u.c.$, that is, about $6.1$nm. This is precisely the critical film thickness of the fcc $\rightarrow$ hcp transition reported in Ref.\cite{Prozenko}, which was obtained using the phenomenological theory proposed in Ref. \cite{Bublik}. The formation of a twin (fcc/hcp) film was observed in our simulation up to a thickness of $30-32~u.c.$ corresponding to about $11-12$ nm. 
\begin{figure}[tbh]
\begin{center}
\resizebox{0.99\columnwidth}{!}{\includegraphics*[angle=-90]{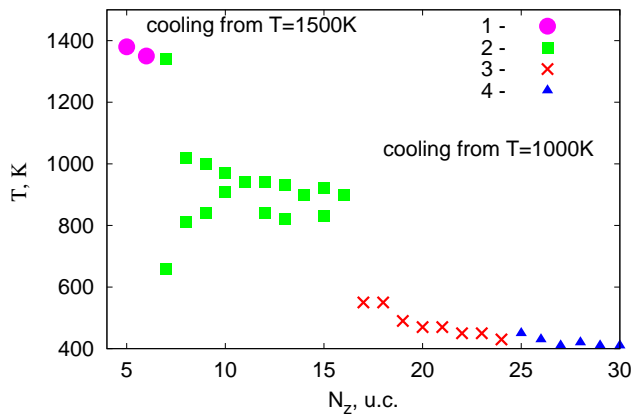}}
\caption{Dependence of the  bcc$\rightarrow$hcp transition temperature on the thickness of bcc (001) Zr film, obtained by "slow cooling". $N_{x,y}= 24 u.c.$. 1 is the one-stage hcp phase, 2 is the "diffuse" transition to the hcp phase, 3 is the fcc/hcp mixture formed by the shear deformation mechanism, 4 is the twin hcp phase formed by martensitic mechanism, with the twin orientation in accordance with long-wave vibrations.
}
\label{Fig4}
\end{center}
\end{figure}

\subsection{Structural transformations on slow cooling}

\begin{figure}[tbh]
\begin{center}
\resizebox{0.45\columnwidth}{!}{\includegraphics*{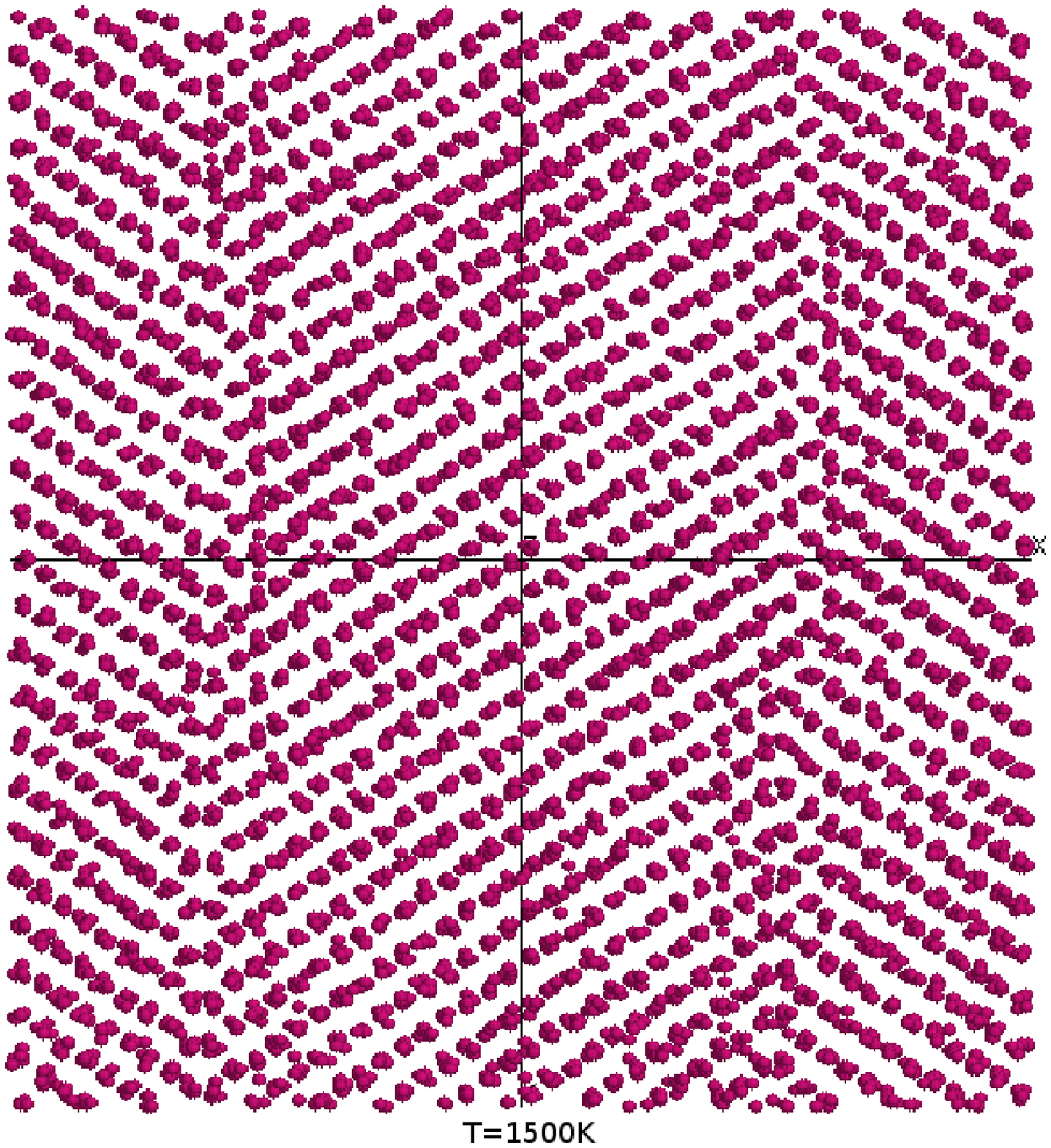}}
\resizebox{0.45\columnwidth}{!}{\includegraphics*{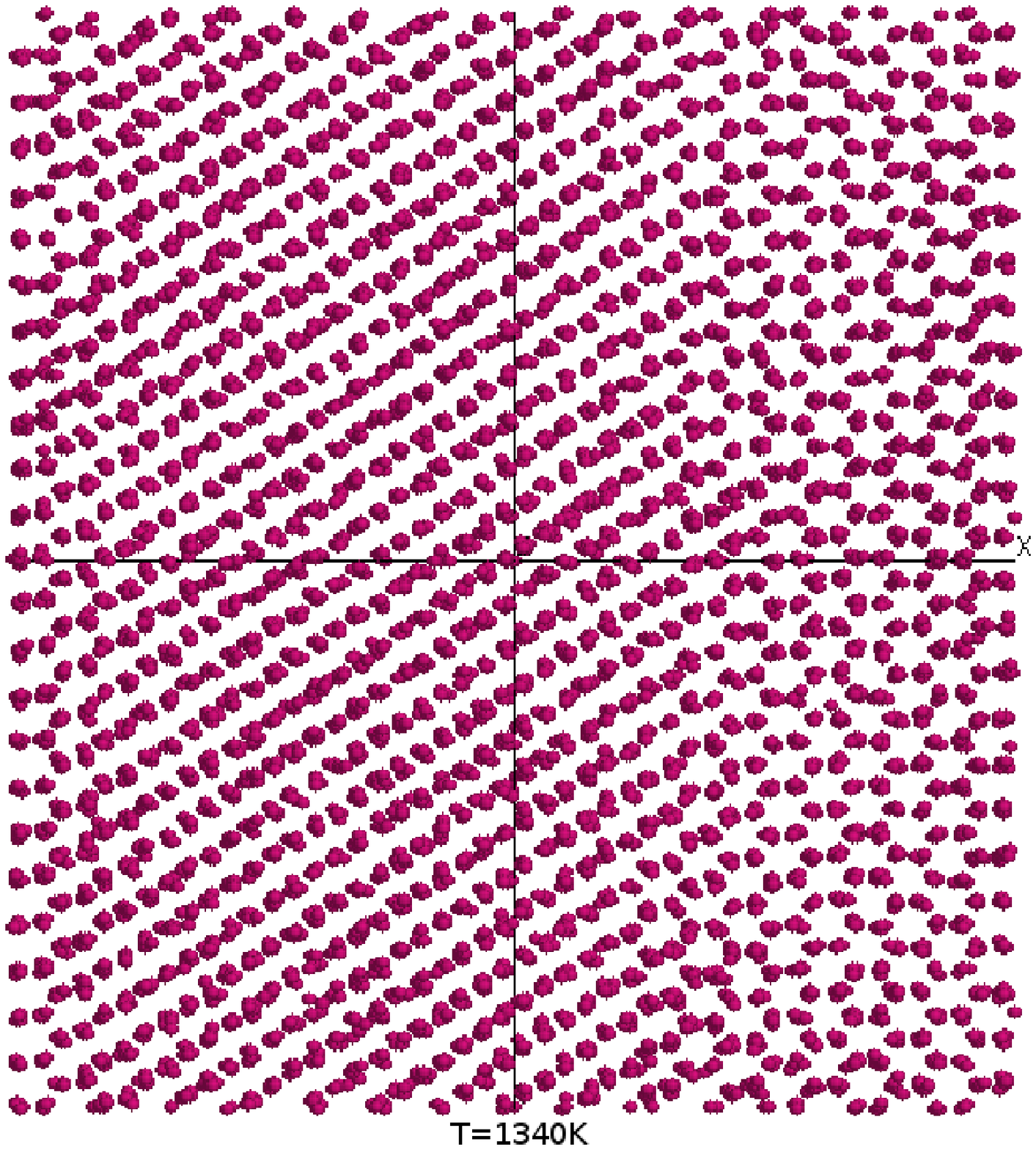}}
\resizebox{0.45\columnwidth}{!}{\includegraphics*{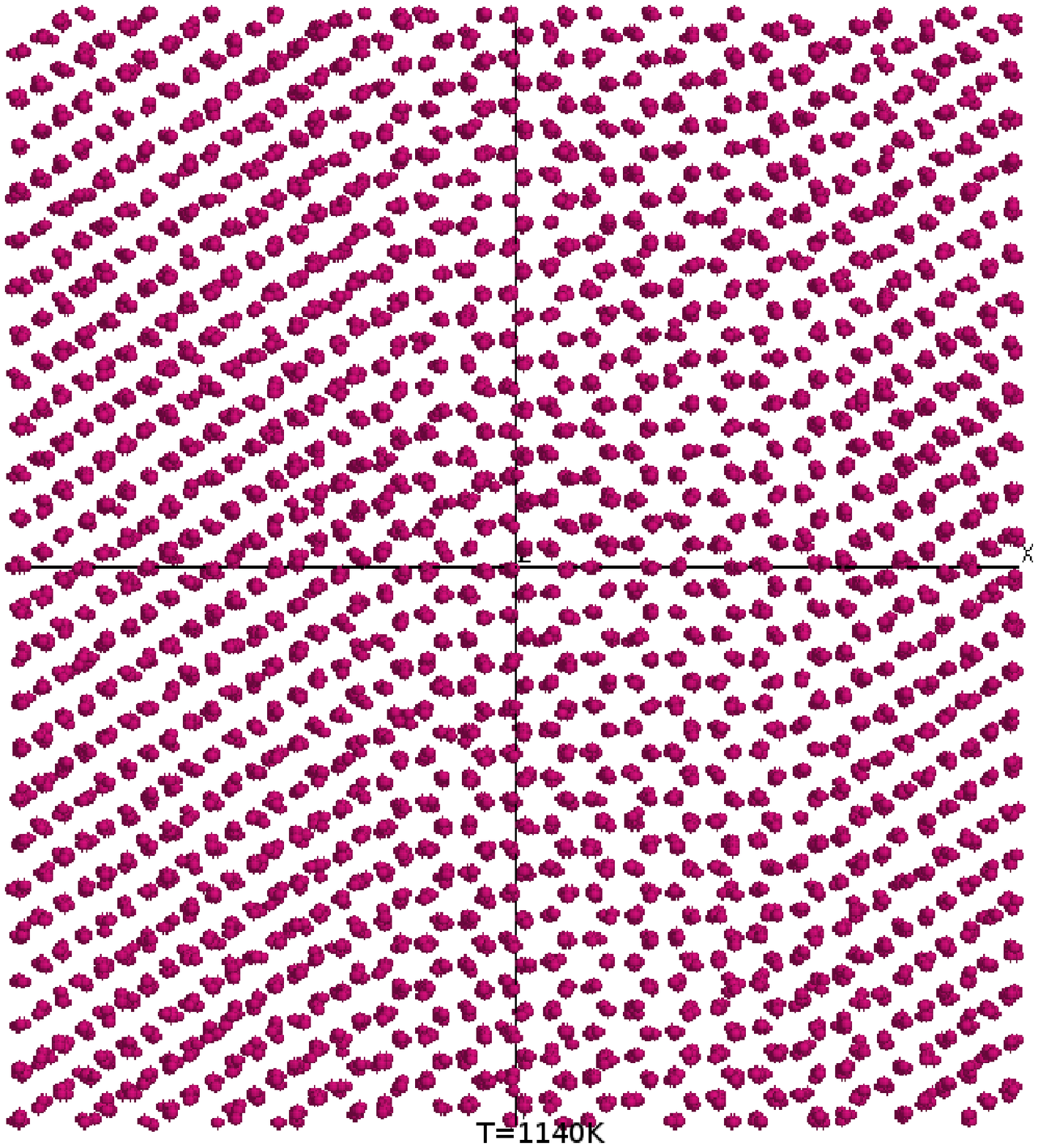}}
\resizebox{0.45\columnwidth}{!}{\includegraphics*{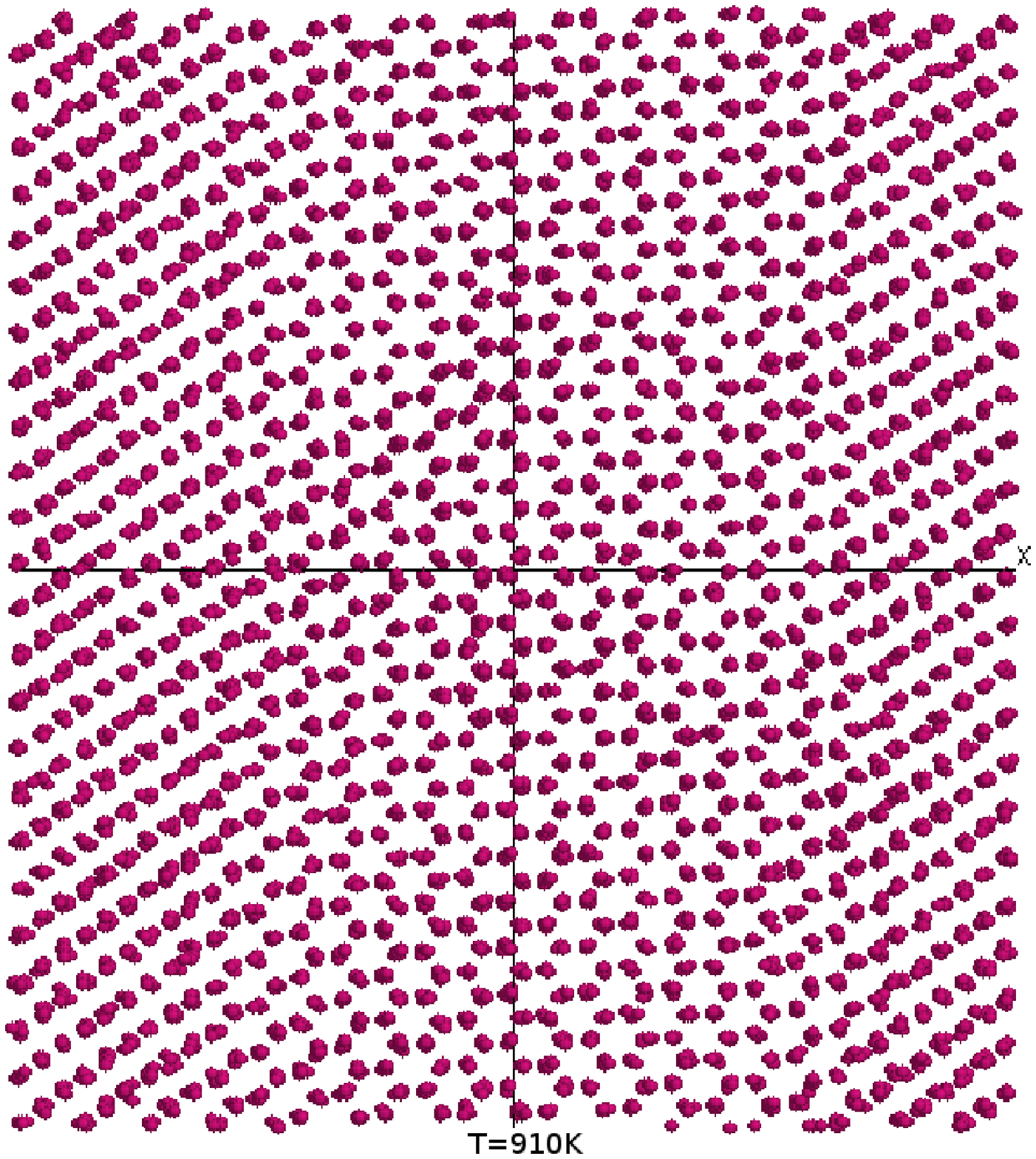}}
\resizebox{0.45\columnwidth}{!}{\includegraphics*{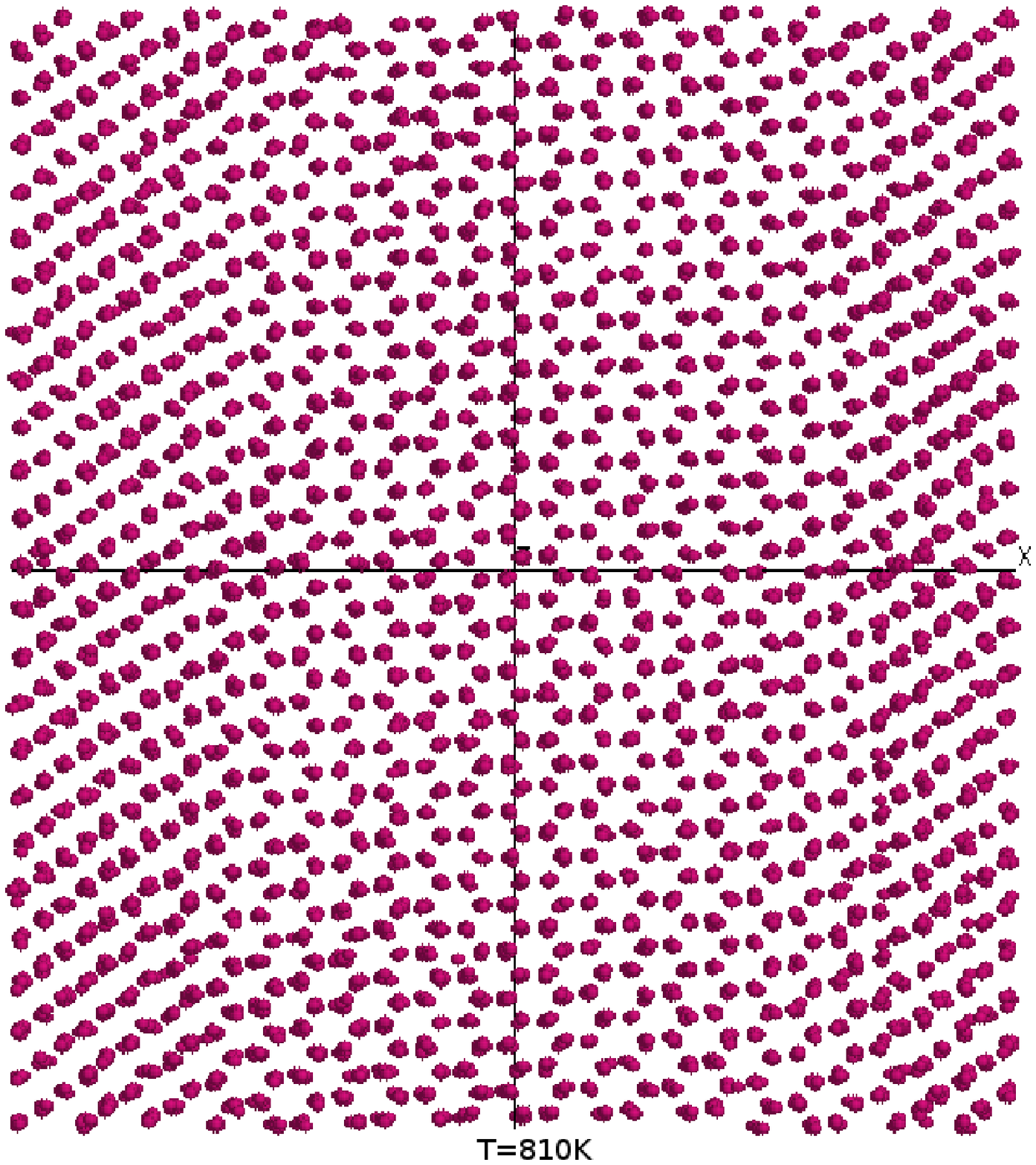}}
\resizebox{0.45\columnwidth}{!}{\includegraphics*{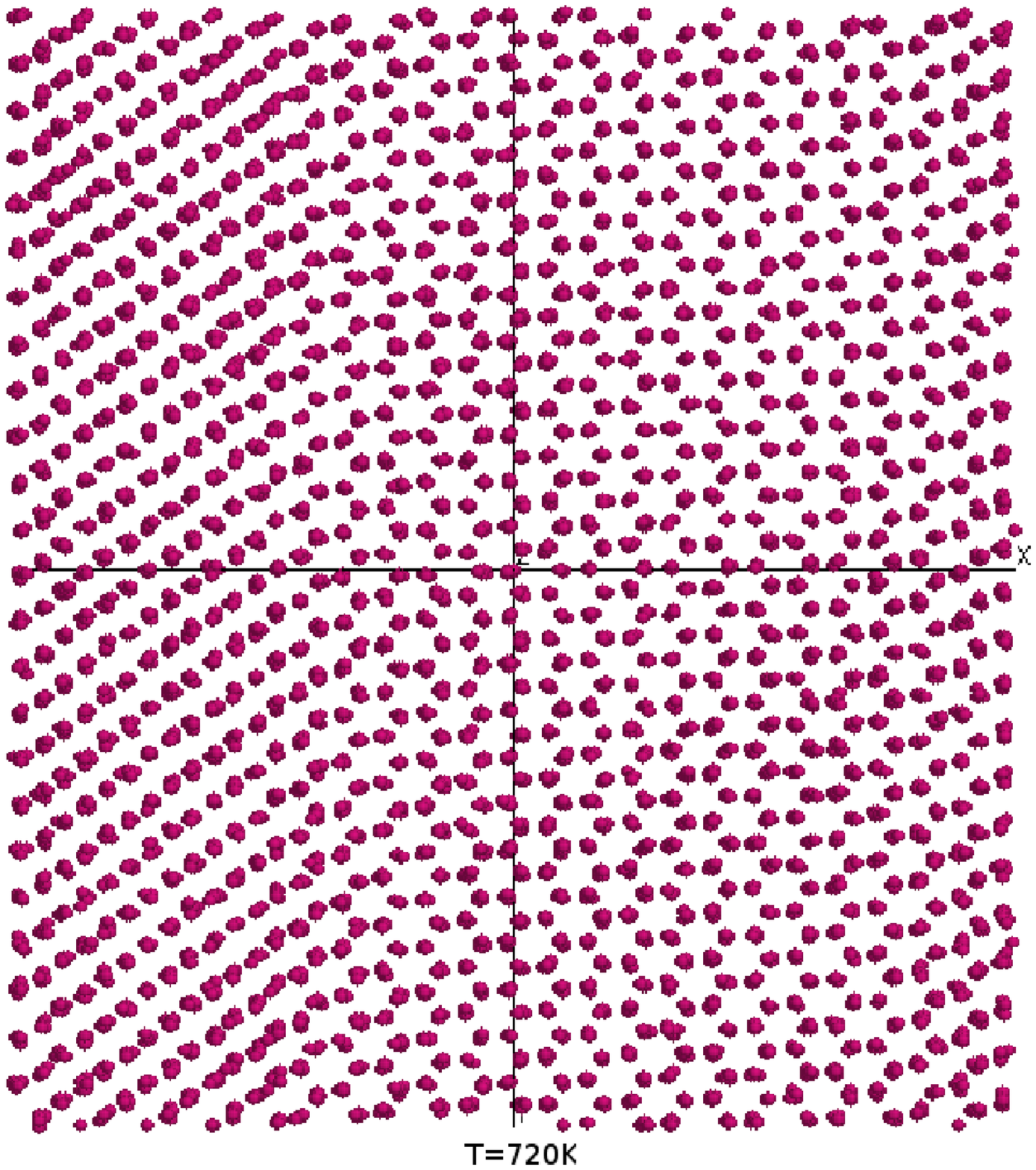}}
\caption{The diffuse transition.
}
\label{Fig5}
\end{center}
\end{figure}

We have also studied the phase transformations in bcc (001) zirconium films by the "slow cooling" technique. Thin films were cooled from a temperature $T=1500 K$, and the thick ones from $T=1000 K$, in both cases with a step $\varDelta T=10 K$ and a holding time $t = 50 ps$ at each new temperature. The calculation results are shown in Fig.\ref{Fig4}. Here the film thickness $N_{z}$ in unit cells is plotted on the horizontal axis, the crystallite dimensions along the x, y axes being the same in all computation variants: $N_{x,y} = 24~u.c.$; the temperature is plotted on the vertical axis.  The temperatures of different structural transformations are marked by different symbols. The films with $N_{z}=5, 6 u.c.$ after reorientation to the bcc (110) phase, completely transform at high temperature into the hcp phase. The temperature of this transition is shown by circles. In thin films with a thickness $N_{z} = 7 - 16 u.c.$ the transition from the bcc to hcp phase does not occur at one temperature value. Instead we observe the so-called "diffuse" transition which is characteristic of martensitic transformations when the phase state changes not abruptly but gradually over a certain temperature range \cite{Malygin-ufn2001}.  In our calculations we observed phase transitions occurring in two stages. The initial and final temperatures of such extended in T transformations are plotted by squares. For instance, in a film with $N_{z} = 7 u.c.$ the transition began at $T = 1340 K$, at that  moment a part of the basic crystallite transformed into the hcp phase, which was accompanied by a small jump in the total energy curve. Only when the temperature was lowered to $T = 660 K$, the remaining part of the crystallite changed to the hcp structure, again with a jump in the total energy curve. The sum of the energies of these two jumps was equal to the value of the energy barrier between the bcc and hcp phases obtained in the case of the one-stage phase transition. The diffuse transition is displayed in Fig.\ref{Fig5}, where a film of dimensions $N_{z}=7~u.c., N_{x,y}=24~u.c.$ is pictured from the [001] direction at different instants of cooling. The corresponding temperatures are indicated under each picture. It is of interest that the crystallite part which has transformed into the hcp phase at the initial temperature, chaotically moves through the crystallite along the [100] direction, slightly increasing in size, and when a critical volume is attained, at the final temperature all the remaining part of the bcc (110) phase abruptly changes to the hcp structure. Note that if the crystallite is held at an intermediate temperature, say, at $T=1000 K$ for 1000 ps (1 million steps), the transition is not completed and the film part transformed to the hcp phase continuously travels through the crystallite without growing in size. A similar effect, i.e., the existence in a wide temperature range of an intermediate configuration of the uncompleted phase transition, was observed in studying the influence of the size of free nanocrystals on the  $\beta\rightarrow\alpha$ transformation in zirconium \cite{our-2009}).
As seen in Fig.\ref{Fig4}, with increasing film thickness the temperature interval between the transition onset and completion decreases . For instance, in a film with $N_{z}=15~u.c.$, the transition starts at $T=920 K$ and is completed at $T=830 K$, while for a thickness $ N_{z}=16~u.c.$ the transformation into the hcp phase occurs immediately in the whole crystallite at a temperature $T=920 K$.

A considerable difference in transition temperature is observed in the figure between the thin and the thick films, which is due to the fact that in thin films the transition proceeds from the reoriented twin bcc (110) phase, while in thick films with $ N_{z}=17~u.c.$ and more it starts from the bcc (001) phase. Moreover, the films with a thickness $N_{z}=17 - 24~u.c.$  transform not in the hcp phase, as might be expected, but in the twin fcc phase. This phase is marked by crosses in Fig.\ref{Fig4}. Note that on cooling these films down to $T=100 K$ with the same procedure their structure does not change, i.e. the twin fcc phase remains stable.  With an increase in the film thickness the temperature of the bcc$\rightarrow$fcc transformation slightly decreases. In the arising twin fcc phase, thin interlayers of hcp phase are observed along the twin boundaries. As in calculations by the "isothermal holding" technique, with increasing film thickness the twin plates grow in number, and consequently increases the hcp portion. And again this increase of the hcp portion occurs not through the conventional martensitic transformation as in bulk materials, but as the result of an increase in the number of twin plates which are shifted relative to each other in the [112] direction. Although the transition takes place in a rather short time interval of 10 - 20 ps, it is realized by shear deformation. The final structure of the film with a thickness $N_{z}=24 u.c.$ after transformation at $T=430 K$ consists almost entirely of a hcp phase being a system of twin plates. The surface of the film obtained has a stepped relief with a step width of 2 or 3 atoms.

\begin{figure}[tbh]
\begin{center}
\resizebox{0.95\columnwidth}{!}{\includegraphics*[angle=-90]{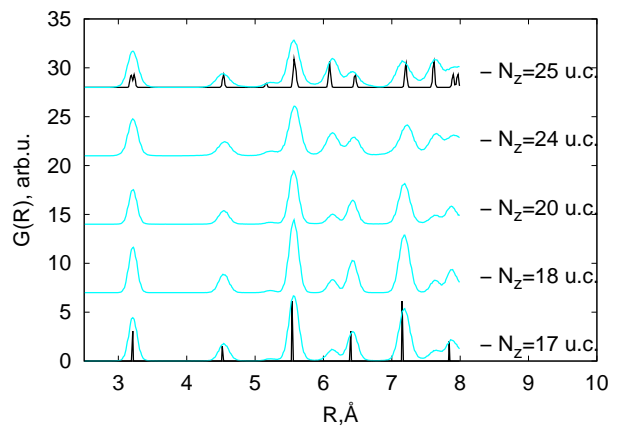}}
\caption{ARDF for the final fcc/hcp mixture in thick films of different thickness.
}
\label{Fig6}
\end{center}
\end{figure}

In Fig.\ref{Fig6} are plotted the atomic radial distribution functions for such mixed (fcc/hcp) phases obtained after structural transformations and then cooled down to T = 100 K, for films of different thickness $N_{z}$. The dark line on bottom plot shows the ARDF for the perfect fcc lattice, and on top plot the ARDF for the hcp structure. It is seen from the figure that as the film thickness grows from $N_{z}=17 u.c.$ to $N_{z}=25 u.c.$, the intensity of peaks at distances 5.15\AA{} and 6.15\AA{} corresponding to the hcp structure gradually increases, and the peak at 6.3 \AA{} (fcc phase) decreases. For the film with $N_{z}=25 u.c.$ the ratio between the peak intensities closely corresponds to the ARDF of hcp lattice.

With further increase in the film thickness to $N_{z}=25 u.c.$ and more, the sequence and mechanism of transformations changes again. In this case the bcc$\rightarrow$hcp transition is realized by the mechanism of cooperative atomic displacements as in bulk zirconium, and the final film structure looks quite different. Instead of a distorted film with stepped surface relief we observe, as a rule, a structure formed by alternating twin hcp regions oriented in accordance with long-wave vibrations arising at given dimensions of the original crystallite (in Fig.\ref{Fig4} the temperatures of the formation of wavy twin hcp structures are marked by triangles). So at "slow cooling" there exists a very narrow thickness range $N_{z}=17 - 23 u.c.$ corresponding to 6.1 - 8.2 nm in which the bcc (001) zirconium film transforms to a nanostructured film of fcc phase. The thickness values for films with a stable twin fcc phase obtained when simulating structural transformations (6.1-8.2 nm) agree with the results of ab initio calculations \cite{Xiong}  of the critical size of zirconium particles with energetically favorable fcc phase (about 6 nm), as well as the experimental data (5-10 nm) \cite{Manna}. 

Thus the change in the parameter $N_{z}$, i.e. the thickness of the bcc zirconium film with (001) surface, leads to changes in the sequence of structural transitions, the mechanism of the bcc $\rightarrow$ hcp transformation, the temperature of phase transitions, and the final phase. Furthermore we have found that also the way of setting the temperature regime substantially affects the process and final structures of phase transformations in films, which was not observed in simulation of the phase transitions in bulk zirconium with cyclic boundary conditions when changing the size of the basic crystallite \cite{our-2012}.

\section{Stability of FCC zirconium films. Elastic moduli}

\begin{figure}[tbh]
\begin{center}
\resizebox{0.95\columnwidth}{!}{\includegraphics*[angle=-90]{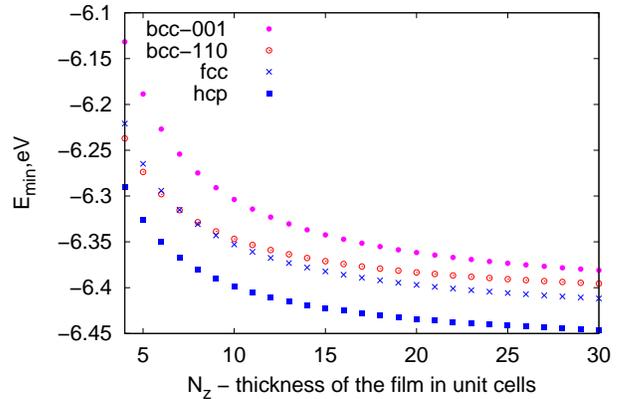}}
\caption{Total energy of different Zr phases as a function of the film thickness at $T=0~K$. 
}
\label{Fig7}
\end{center}
\end{figure}

In paper \cite{Bublik}  it has been shown that whereas in bulk samples the bcc lattice is stable, in thin films the fcc structure may form, since a more closely packed phase has a lower surface energy.
As stated above, in zirconium nanofilms a stable fcc phase is realized in a certain temperature range. Calculations of the total energy in the ground state show that fcc zirconium has a higher energy than the hcp phase but lower than the bcc lattice.  At atmospheric pressure in bulk zirconium there exist only two phases: hcp at low temperatures and bcc at high temperatures. As shown in Ref.\cite{Trub}, stabilization of the high-temperature bcc phase is determined by the contributions to the free energy from strongly anharmonic lattice vibrations with wave vector ${\bf k}=2/3(111)$  of the Brillouin zone. 
We do not know any $\textit{ab initio}$ calculations of the ground-state energy in bcc and fcc zirconium films. The total energy of Zr films calculated as a function of thickness using the chosen EAM potential at $T=0~K$ is displayed in Fig.\ref{Fig7}. Based on these data one can estimate the energy difference $\Delta E_{bcc-fcc} = E_{bcc} - E_{fcc}$ at $T=0~K$ for zirconium films and compare it with that for bulk samples. The results obtained show that the presence of the surface does not change the sign of $\Delta E$ relative to the bulk calculation: for the surface $\Delta E_{bcc-fcc}^{surf}=50$meV, and for the bulk $\Delta E_{bcc-fcc}^{vol}=21$meV.

It should be noted that with increasing temperature the difference in energy between the bcc and fcc phases decreases : $\Delta E_{bcc-fcc}^{surf}=37$meV at $T=1100$K. (At $Т=1100~К$ the fcc phase is metastable, see \cite{our-2012}). Keeping in mind that the hcp zirconium films at $T=0~K$ are energetically preferable ($\Delta E_{fcc-hcp}^{surf}=21$meV), one can suppose that the fcc phase stability is also related to the peculiarities of lattice vibrations. So the formation of a stable fcc phase in Zr films may be connected with the contributions to the free energy not only from the surface states but also from those determined by the lattice anharmonicity. 

\begin{table*}[!tbh]
\begin{center}
\begin{tabular}{|c|c|c|c|c|c|c|c|c|}
\hline
phase &method & ${a}_{0}$ & $V_{0}$ &B & C' & $C_{11}$  & $C_{12}$ & $C_{44}$ \\
\cline{1-9}
&\bf{ our calc }& \bf{3.563}& \bf{22.616 } &\bf{106.5} & \bf{- 9.2} & \bf{95.5} & \bf{112 }& \bf{41.9}\\
\cline{2-9}
&WASP \cite{Mendelev-Ack} & 3.57 & 22.75  & 89.3  & - 5.5  & 82  & 93 & 29 \\
\cline{2-9}
&LDA \cite{Ji-2003} & 3.47 & 20.87  & 93.2 & 11.65  & 108.7  & 85.4 & 10.6 \\
\cline{2-9}
bcc-bulk&FPLMTO \cite{Mahdi} & 3.568 & -  & 90.4  & -  & - & - & - \\
\cline{2-9}
&EAM1 \cite{Kum} & 3.586 & -  & 92.4  & -  & - & - & - \\
\cline{2-9}
&EXP \cite{Phonon-bcc}  & 3.574 & 22.83  & 96.67  &  6  & 104 & 93 & 38\\
\cline{1-9}
&\textbf{our calc}  & \textbf{4.544 }&\textbf{ 23.463} & \textbf{87.7} & \textbf{16.53 }& \textbf{109.74} & \textbf{76.68} & \textbf{64.33}\\
\cline{2-9}
&FPLAPW \cite{Aguayo} & 4.52& 23.1 & 91 & 21 &  119  & 77  & 53\\
\cline{2-9}
&EAM1 \cite{Kum} & 4.542&  -  & 87.4 & - & -  & -  & -\\
\cline{2-9}
fcc-bulk&EAM2 \cite{Li}  & 4.53&  -  & 94 & - & 130  & 76  & 65\\
\cline{2-9}
&CASTEP \cite{M} & 4.52&  -  & 90.18 & 15.495 & 110.84  & 79.85  & 48.24\\
\cline{1-9}
fcc-film&\textbf{ our calc } & \textbf{4.537} & \textbf{23.35} & \textbf{ 91.2 }& \textbf{ 15.35 } & \textbf{111.68} &\textbf{ 81 }& \textbf{63.47}\\
\cline{2-9}
&EXP\cite{Chopra} & 4.61&  -  & - & - & -  & -  & -\\
\cline{2-9}
&EXP\cite{Hill} & 4.63&  -  & - & - & -  & -  & -\\
\cline{2-9}
&EXP\cite{Ji-2003} & 4.61&  -  & - & - & -  & -  & -\\
\cline{2-9}
&EXP\cite{Prozenko} & 5.04&  -  & - & - & -  & -  & -\\
\cline{1-9}
 \hline
\end{tabular}

\end{center} 
\label{table:Tabl1}

\caption{Table 1. Lattice parameters  $ a_{0}$  (\AA),   equilibrium  volume $V_{0}$(\AA$^{3}$/atom), bulk modulus($B$), shear modulus(C') and elastic constants (GPa) for Zr in the bcc and fcc phases.}

 \end{table*}

The stability of fcc zirconium is also demonstrated by the elastic moduli calculated by various methods. For comparison in Table 1 are listed the data for bcc and fcc zirconium structures obtained in our and other papers, including the experimental values of the lattice parameter of fcc films. 
In the second column of the Table are indicated the abbreviated names of the calculation methods used in the corresponding papers, namely, the ab initio calculations WASP, FPLMTO, FPLAPW, CASTEP, LDA, and the MD ones using the potentials constructed in the embedded-atom model: EAM1 and EAM2.
\begin{figure}[tbh]
\begin{center}
\resizebox{0.95\columnwidth}{!}{\includegraphics*[angle=-90]{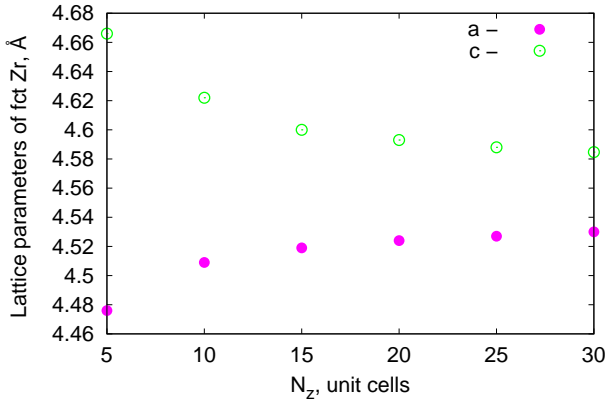}}
\caption{Variation of the lattice parameters ($a$ and $c$) with the fcc film thickness at $T=300~K$.}
\label{Fig8}
\end{center}
\end{figure}

Note that in the LDA calculations \cite{Ji-2003} the bcc lattice was found to be stable at $T=0~K$. Our calculations show that the bcc lattice has a negative shear modulus  $C'$ at zero temperature, i.e. at low temperatures the bcc lattice is unstable as it must. The experimental data for bcc zirconium \cite{Phonon-bcc} were obtained at $T=915~K$.
The lattice parameter for both the bulk and film samples was determined from the total energy minimum in a non-relaxed crystallite at $T=0~K$. The lattice parameter for the fcc film was obtained for a non-relaxed film with a thickness  $N_{z}=10~u.c.$

It should be mentioned here that in films the lattice parameter depends on both the film thickness and the temperature. For example, in the same fcc film of thickness $N_{z}=10~u.c.$ as the temperature grows from $T=100~K$ to $500~K$, the volume per atom changes from $23.42$\AA$^{3}$ to $23.56$\AA$^{3}$, while in a film with $N_{z}=17 u.c.$ under the same temperature variation the volume changes from $23.45$\AA$^{3}$ to $23.57$\AA$^{3}$.
It is seen from the Table that the lattice parameters of the experimentally obtained fcc zirconium films differ from one another, as the films in these works were prepared in different ways, under different conditions, and on different substrates which are known to substantially affect the realization and stability of a structure, and consequently its lattice parameter.
Moreover, as shown by our calculations, the bcc and fcc films lose the cubic symmetry and become tetragonal (bct and fct, respectively), the lattice stretches along the $z$ direction with free boundary conditions. A tetragonal distortion of the film was also found experimentally \cite{Ji-2003}. 
Figure \ref{Fig8} displays the variation of the lattice parameters $c$ (along the $z$ direction), and $a$ (along the $x,y$ 
directions) with increasing the film thickness. The calculations were performed for crystallites relaxed during 100 ps at $T=300~K$, and then the atomic positions were averaged over the 50 ps range for films of different thickness. As seen from the figure, with an increase in the film thickness the parameter $c$ diminishes, and the parameter $a$ grows, the ratio $c/a$ being equal to 1.042 for the film with $N_{z}=5~u.c.$, and 1.011 for the film with $N_{z}=30~u.c.$; that is, the thinner the film, the greater is the tetragonal distortion of the lattice.
The calculations of the bulk modulus $B$, and shear constants $C'$ and $C_{44}$ were performed by the formulas:

\begin{equation}
B=\frac{ a_{0}^{2}}{9V} \frac{d^{2}E}{da^{2}}
\end{equation}
\begin{equation}
C'= \frac{a_{0}^{2}}{4V } \frac{d^{2}E}{da^{2}}
\end{equation}

where $a$ is the lattice parameter, $a_{0}$ is the equilibrium lattice parameter, $V$ is the volume per atom, $E$ is the total energy per atom as a function of the lattice parameter $a$.
The calculation of $C_{44}$ was also carried out by formula (2) but for the (110) orientation of the fcc lattice, i.e. with the $x$ axis along the [110] direction, and the  $y$ axis along the direction [1-10]. Since the calculation was performed for non-relaxed crystallites, the relations between the bulk and the elastic moduli correspond to the cubic symmetry:

\begin{equation}
B=\frac{C_{11}+2 C{12}}{3}
\end{equation}
\begin{equation}
C'= \frac{C_{11}-C_{12}}{2} 
\end{equation}

\begin{figure}[tbh]
\begin{center}
\resizebox{0.85\columnwidth}{!}{\includegraphics*[angle=-90]{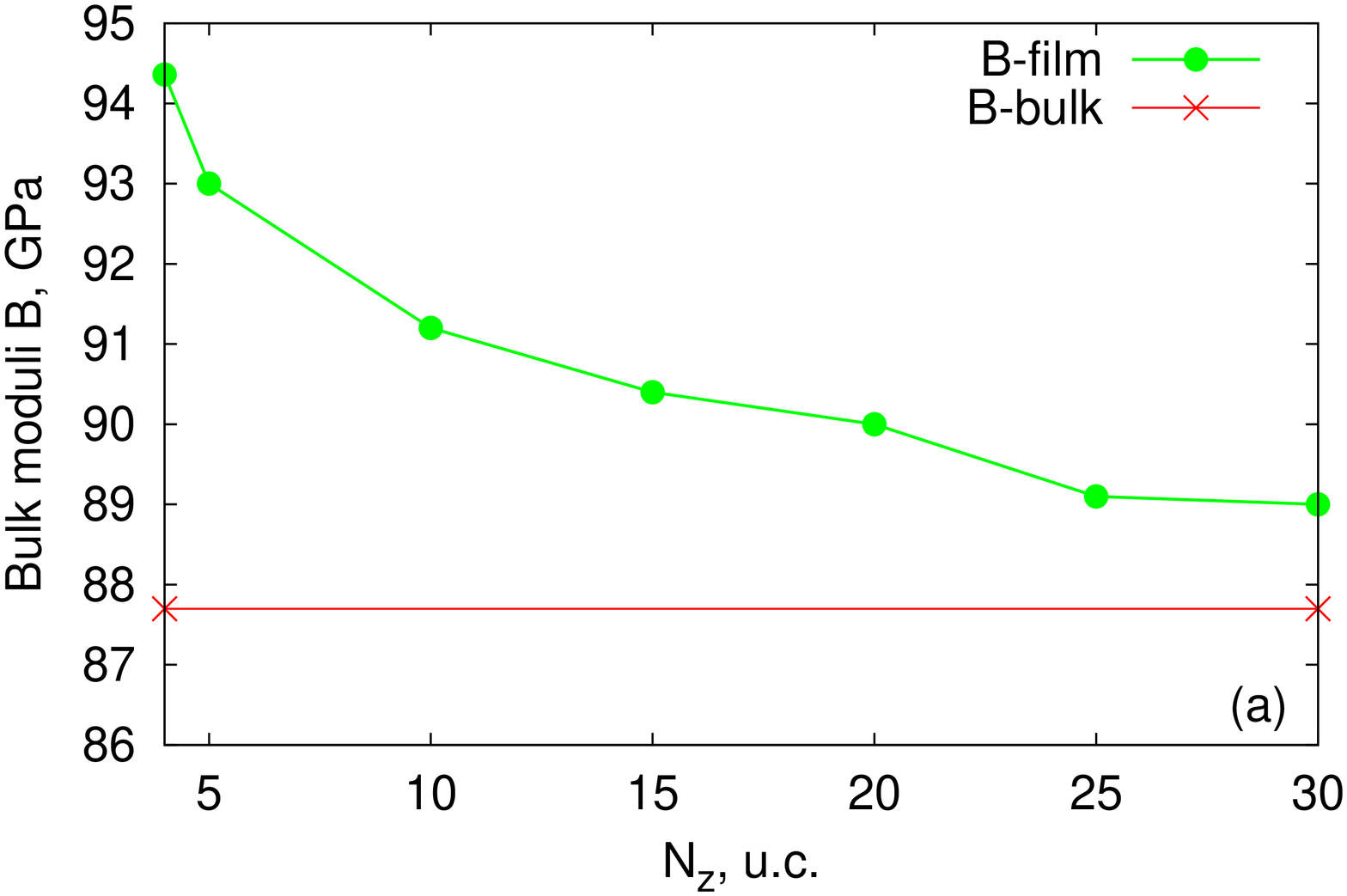}}
\resizebox{0.85\columnwidth}{!}{\includegraphics*[angle=-90]{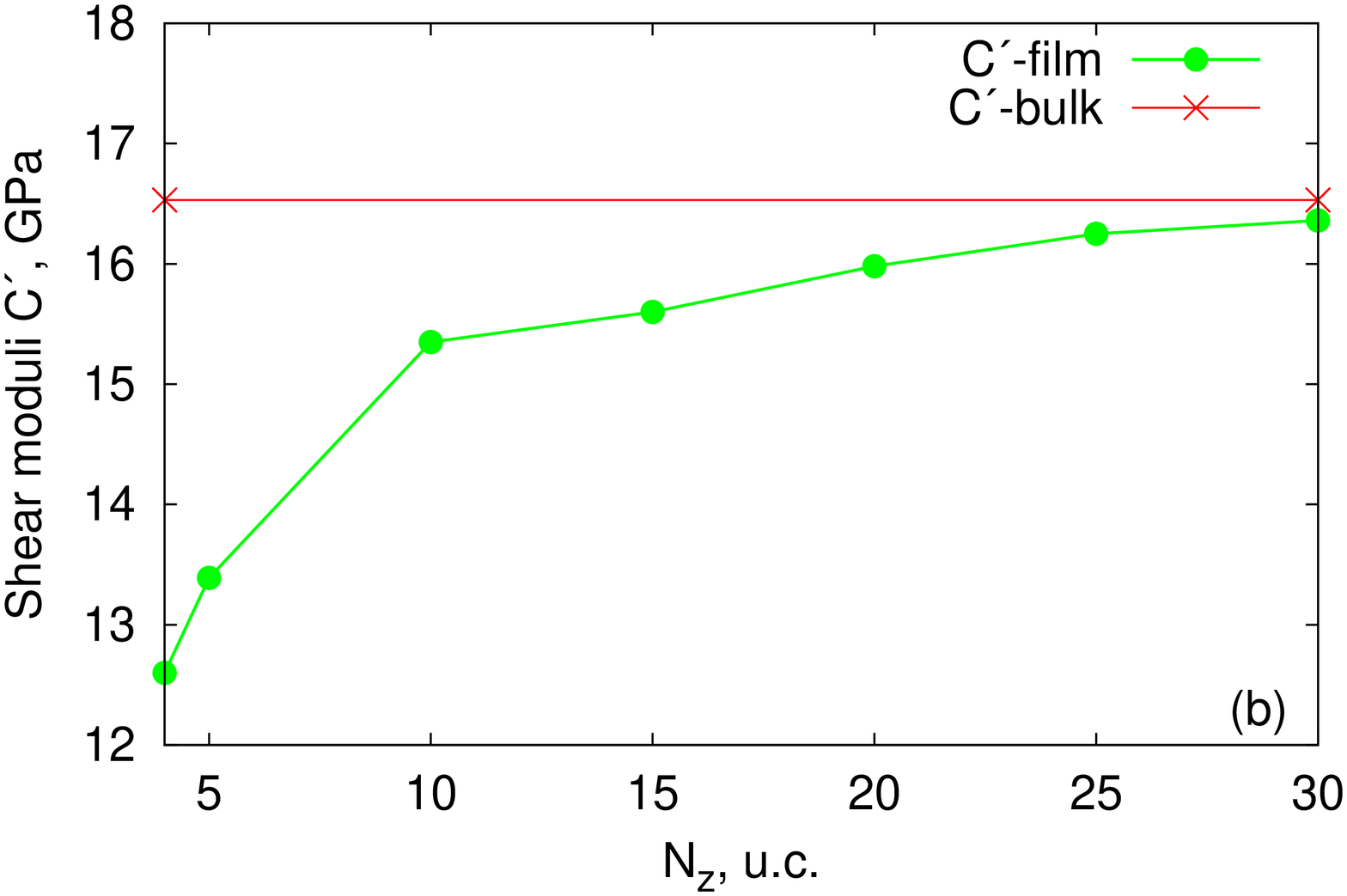}}
\resizebox{0.85\columnwidth}{!}{\includegraphics*[angle=-90]{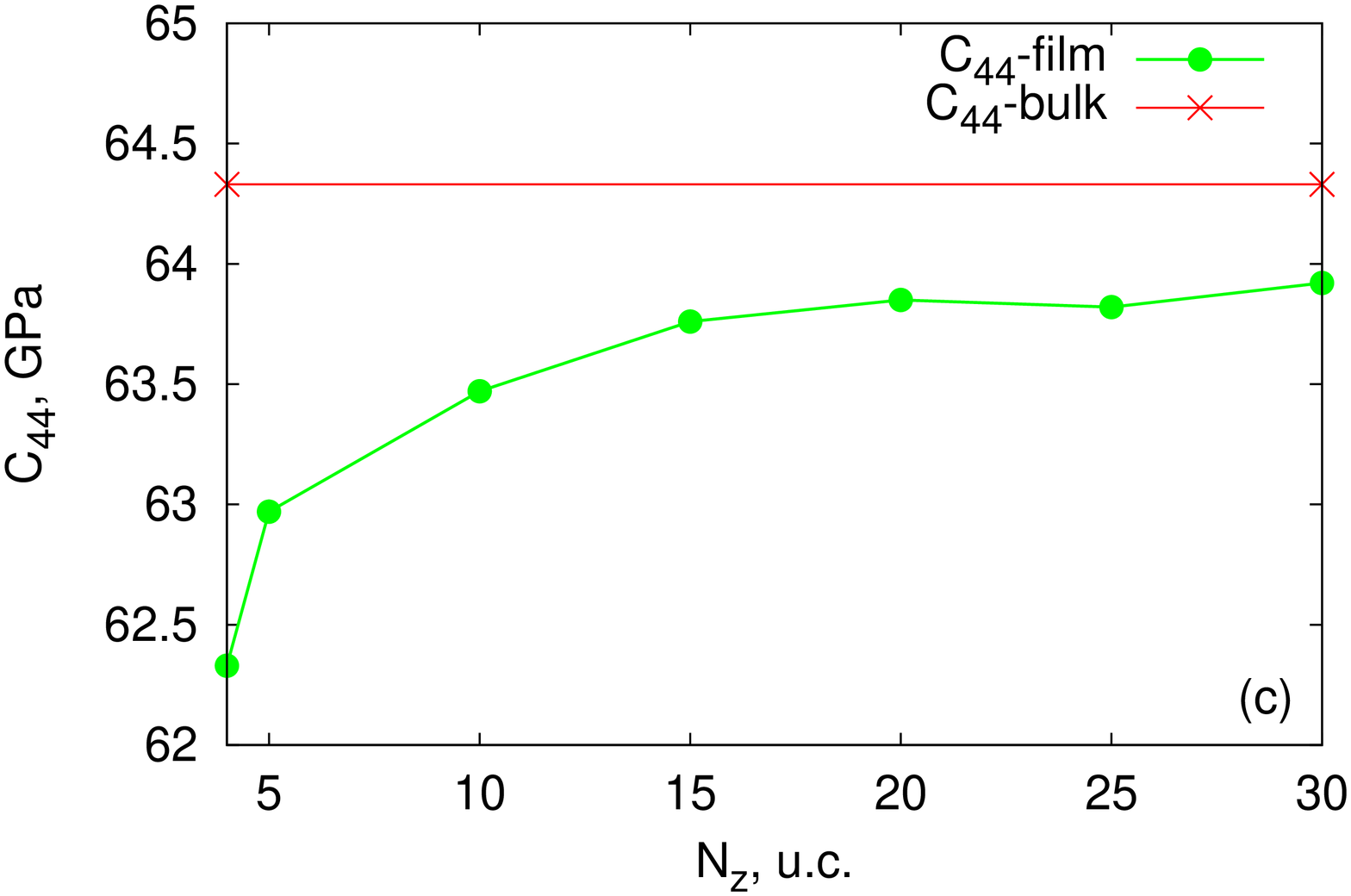}}
\caption{Dependence of the bulk modulus (a), shear modulus (b), and $C_{44}$ (c) on the non-relaxed fcc Zr film thickness.}
\label{Fig9}
\end{center}
\end{figure}

In addition, we have calculated the dependence of the elastic moduli on the thickness of a \textit{non-relaxed} fcc zirconium film. Figure \ref{Fig9} presents the size dependences of the bulk modulus $B$, and shear moduli $C'$ and $C_{44}$. The abscissa is the film thickness $N_{z}$ in unit cells, the straight line on the plots corresponds to the bulk value of the given quantity. One can see that with increasing film thickness the moduli behave in different ways: the bulk modulus $B$ diminishes, while the moduli $C'$ and $C_{44}$ increase, approaching their bulk values. All the moduli values are positive, which points to the stability of fcc films at low temperatures. 

\section{Conclusion}

Thus the study performed has evidenced the existence of a temperature range (depending on the film thickness) in which a stable fcc phase may form in nanofilms of pure zirconium. The magnitude of tetragonal distortion in this phase depends on the film thickness, and ranges from 4\% to 1\% as $N_z$ increases from 5 to 30. In films thicker than 6.1 nm the fcc phase has a twin structure, and a hcp structure forms at twin boundaries. The portion of hcp phase increases with the film thickness. At $N_z > 32~u.c.$  (about 12 nm) the fcc phase is not observed. An analysis of the total energy of different phases in Zr films and of the elastic moduli of the bcc and fcc phases at $T=0~K$ suggests that the fcc phase stability at finite temperatures is determined by the lattice vibrations.  

Studying the morphology of the structures obtained after transformations in bcc zirconium films has shown that the mechanisms of structural transformations substantially depend on the film thickness and orientation. For bcc (001) zirconium films a limiting thickness of 6.1 nm was found, at which the mechanism, sequence, and temperature of the structural transitions, as well as the final phase do change.  In thin films the morphology of the structure observed after transformations in bcc (001) films essentially depends also on the film nanostructure (in our calculation it was modeled by the basic crystallite dimensions in the MD simulation along the x, y axes). In particular, an orthorhombic $\alpha ''$  phase was shown to form for odd $N_{x,y}<20$.

In thin (up to 6.1 nm) films under slow cooling a "diffuse" phase transition is observed. The temperature interval between the onset and the completion of structural transition narrows with increasing film thickness, and the temperature of the onset of structural transformation decreases.  

\begin{acknowledgments}
The authors acknowledge  the partial support from the  RFBR Grant $ N^{o}$ 10-02-96034-r-ural-a, 
Grants  of the Presidium of the Russian Academy of Sciences $ N^{o}$ 12-P-12-2010 
and Ural Branch of Russian Academy of Sciences $ N^{o}$ 12-Y-2-1023
\end{acknowledgments}

\end{document}